\documentclass{emulateapj}

\begin{document}

\title{Hemispherical power asymmetry in the
  three-year\\ Wilkinson Microwave Anisotropy Probe sky maps}

\author{H.\ K.\ Eriksen\altaffilmark{1,2,3,4}, A. J.
  Banday\altaffilmark{5}, K.\ M.\ G\'orski\altaffilmark{3,4,6}, F. K.
  Hansen\altaffilmark{1,2} and P.\ B.\ Lilje\altaffilmark{1,2}}

\email{h.k.k.eriksen@astro.uio.no}

\altaffiltext{1}{Institute of Theoretical Astrophysics, University of
  Oslo, P.O.\ Box 1029 Blindern, N-0315 Oslo, Norway}

\altaffiltext{2}{Centre of Mathematics for Applications, University of
  Oslo, P.O.\ Box 1053 Blindern, N-0316 Oslo, Norway}

\altaffiltext{3}{Jet Propulsion Laboratory, 4800 Oak Grove Drive,
  Pasadena CA 91109}

\altaffiltext{4}{California Institute of Technology, Pasadena, CA
  91125}

\altaffiltext{5}{Max-Planck-Institut f\"ur Astrophysik,
  Karl-Schwarzschild-Str.\ 1, Postfach 1317, D-85741 Garching bei
  M\"unchen, Germany}

\altaffiltext{6}{Warsaw University Observatory, Aleje Ujazdowskie 4,
  00-478 Warszawa, Poland}


\begin{abstract}
  We consider the issue of hemispherical power asymmetry in the
  three-year WMAP data, adopting a previously introduced modulation
  framework. Computing both frequentist probabilities and Bayesian
  evidences, we find that the model consisiting of an isotropic CMB
  sky modulated by a dipole field, gives a substantially better fit to
  the observations than the purely isotropic model, even when
  accounting for the larger prior volume. For the ILC map, the
  Bayesian log-evidence difference is $\sim1.8$ in favour of the
  modulated model, and the raw improvement in maximum log-likelihood
  is 6.1. The best-fit modulation dipole axis points toward $(l,b) =
  (225^{\circ},-27^{\circ})$, and the modulation amplitude is 0.114,
  in excellent agreement with the results from the first-year
  analyses. The frequentist probability of obtaining such a high
  modulation amplitude in an isotropic universe is $\sim1$\%.  These
  results are not sensitive to data set or sky cut.  Thus, the
  statistical evidence for a power asymmetry anomaly is both
  substantial and robust, although not decisive, for the currently
  available data.  Increased sky coverage through better foreground
  handling and full-sky and high-sensitivity polarization maps may
  shed further light on this issue.
\end{abstract}

\section{Introduction}
\label{sec:introduction}


While the first-year results from the Wilkinson Microwave Anisotropy
Probe (WMAP) experiment \citep{bennett:2003} overall clearly supported
the currently popular inflationary cosmological model, describing a
flat, isotropic and homogeneous universe seeded by Gaussian and
adiabatic fluctuations, a disturbing number of unexpected anomalies on
large scales were reported shortly after the public data
release. Perhaps the three most important ones were: 1) alignments and
symmetry features among low $\ell$ multipoles \citep{de
Oliveira-Costa:2004, eriksen:2004a}, 2) an apparent asymmetry in the
distribution of fluctuation power in two opposing hemispheres
\citep{eriksen:2004b, hansen:2004}, and 3) a peculiar cold spot in the
southern hemisphere \citep{vielva:2004, cruz:2005}. All of these
features were subsequently studied extensively by independent groups,
and all remain unresolved to the present day.

In March 2006 the three-year WMAP results were released, prompting
researchers to revisit the anomalies detected in the first-year data
\citep{bridges:2006, copi:2006, jaffe:2006, land:2006,
martinez-gonzalez:2006}. Of course, considering that already the
first-year data were strongly signal-dominated on the scales of
interest, it should come as no surprise that most of these analyses
concluded with similar results as for the previous data, although
different foreground handling could affect some results.

The WMAP team paid particular attention to the question of large-scale
power asymmetry in their analyses \citep{hinshaw:2007, spergel:2007a}.
Specifically, in an early version of their paper,
\citet{spergel:2007a} approached the problem from a semi-Bayesian
point of view, by defining a parametric model consisting of an
isotropic and Gaussian CMB field modulated by a large-scale function.
The power asymmetry anomaly was then addressed by a dipolar modulation
field, and the low-$\ell$ alignment anomalies were studied with a
quadrupole modulation field. However, due to several issues with this
early analysis, several of which were first addressed by the present
paper, the authors decided to remove the corresponding section from
the final version of their paper \citep{spergel:2007b}.  One example
is simple marginalization over non-cosmological monopoles and dipoles
components, which was first done by \citet{gordon:2007} in an
otherwise identical analysis. A second example was the limited
harmonic range considered by \citet{spergel:2007a}. Thus, we present
in this Letter the first complete modulation analysis that covers the
full range of angular scales presented by \citet{eriksen:2004b}, and
that takes into account all known sources of systematics, such as
monopole/dipole and foreground marginalization. We also present the
first proper computation of the Bayesian evidence for the modulated
model.

Following the first report of the power asymmetry, much effort has
been spent by theorists on providing possible physical explanations.
Examples range from those questioning the very fundamentals of physics
and cosmology (e.g., introducing intrinsically inhomogeneous
cosmologies -- Moffat 2005 and Jaffe et al.\ 2005; violation of Lorenz
invariance -- Kanno \& Soda 2006; or violation of rotational
invariance in the very early universe -- Ackermann, Carroll \& Wise
2007) to those essentially considering special cases of established
physics (e.g., second-order gravitational effects from local
inhomogeneities -- Tomita 2005; the presence of local voids -- Inoue
\& Silk 2006; spontaneous isotropy breaking from non-linear response
to long-wavelength density fluctuations -- Gordon et al.\
2005). 

\section{Algorithms}
\label{sec:analysis}

We now outline the methods used for the analyses presented in the
following sections. 

\subsection{Data model and likelihood}

We model the CMB temperature sky maps as
\begin{equation}
  \mathbf{d}(\hat{\mathbf{n}}) = \mathbf{s}(\hat{\mathbf{n}}) [1 + f(\hat{\mathbf{n}})]
  + \mathbf{n}(\hat{\mathbf{n}}),
\end{equation}
where $\mathbf{s}(\hat{\mathbf{n}})$ is a statistically isotropic and
Gaussian random field with power spectrum $C_{\ell}$, $f(\hat{\mathbf{n}})$
is a dipole modulation field with amplitude less than unity, and
$\mathbf{n}(\hat{\mathbf{n}})$ is instrumental noise. Thus, the modulated signal
component is an anisotropic, but still Gaussian, random field, and
therefore has a covariance matrix given by
\begin{equation}
\tilde{\mathbf{S}}(\hat{\mathbf{n}}, \hat{\mathbf{m}}) = [1+f(\hat{\mathbf{n}})]
\mathbf{S}(\hat{\mathbf{n}}, \hat{\mathbf{m}})[1+f(\hat{\mathbf{m}})],
\end{equation}
where
\begin{equation}
\mathbf{S}(\hat{\mathbf{n}}, \hat{\mathbf{m}}) = \frac{1}{4\pi}\sum_{\ell}
(2\ell+1) C_{\ell} P_{\ell}(\hat{\mathbf{n}}\cdot\hat{\mathbf{m}}).
\end{equation}

Taking into account instrumental noise and possible foreground
contamination, the full covariance matrix is
\begin{equation}
\mathbf{C}(\hat{\mathbf{n}}, \hat{\mathbf{m}}) =
\tilde{\mathbf{S}}(\hat{\mathbf{n}}, \hat{\mathbf{m}}) + \mathbf{N} + \mathbf{F}.
\end{equation}
The noise and foreground covariance matrices depend on the data
processing, and will be described in greater detail in \S \ref{sec:data}. 
With these definitions ready at hand, the log-likelihood is given by
\begin{equation}
-2\log \mathcal{L} = \mathbf{d}^T \mathbf{C}^{-1} \mathbf{d} + \log |\mathbf{C}|,
\end{equation}
up to an irrelevant constant. 

\subsection{Posterior distributions and choice of parameters}

The posterior distribution $P(\theta|\mathbf{d})$ is a primary goal of
any Bayesian analysis, $\theta$ being the set of all free parameters
in the model. For the model defined above, the free parameters can be
divided into two groups, namely those describing the isotropic CMB
covariance matrix or $C_{\ell}$, and those describing the modulation
field. Both may be parametrized in a number of different ways, and
these choices may affect the outcome of the analysis through different
prior definitions.

First, for the isotropic CMB component, we choose to parametrize the
power spectrum in terms of a simple two-parameter model with free
amplitude $q$ and tilt $n$,
\begin{equation}
C_{\ell} = q \left(\frac{\ell}{\ell_0}\right)^{n} C_{\ell}^{\textrm{fid}}.
\end{equation}
Here $\ell_0$ is a pivot multipole and $C_{\ell}^{\textrm{fid}}$ is a
fiducial model, in the following chosen to be the best-fit power law
spectrum of \citet{hinshaw:2007}. Second, the modulation field
$f(\hat{n})$ is parametrized in terms of a direction
$\hat{\mathbf{p}}$ and an overall amplitude $A$,
\begin{equation}
f(\hat{\mathbf{n}}) = A \, \hat{\mathbf{n}} \cdot \hat{\mathbf{p}}.
\end{equation}

We use flat priors on all parameters in this paper; the modulation
axis is uniform over the sphere, and the amplitude is restricted to $A
\le 0.3$. The power spectrum parameters are restricted to $0.5 \le q
\le 1.5$ and $-0.5 \le n \le 0.5$. These choices are sufficiently
generous to include all non-zero parts of the likelihood.

The posterior distribution,
\begin{equation} 
P(q, n, A, \hat{\mathbf{p}} | \mathbf{d}) \propto \mathcal{L}(q, n, A,
  \hat{\mathbf{p}}) P(q, n, A, \hat{\mathbf{p}}),
\end{equation}
is then mapped out using a standard Markov Chain Monte Carlo
technique. We use a Gaussian proposal density for $q$, $n$, and $A$,
and an Euler-matrix based, uniform proposal density for
$\hat{\mathbf{p}}$.

\subsection{Bayesian evidence and nested sampling}

In a Bayesian analysis, one is not only interested in the set of
best-fit parameter values, but also in the relative probability of
competing models. The most direct way of measuring this is through the
Bayesian evidence,
\begin{equation}
E \equiv P(\mathbf{d}|H) = \int P(\mathbf{d}|\theta, H) P(\theta|H) d\theta,
\end{equation}
which is simply the average likelihood over the prior volume.
Typically, one computes this quantity for two competing models, $H_0$
and $H_1$, and considers the difference $\Delta \log E = \log E_1 -
\log E_0$. If $\Delta \log E > 1$, the evidence for $H_1$ is
considered substantial; if $\Delta \log E > 2.5$, it is considered
strong.

Traditionally, computation of evidences has been a computational
challenge. However, \citet{mukherjee:2006} introduced a method called
``nested sampling'', proposed by \citet{skilling:2004}, to the
cosmological community that allows for accurate estimation of the
evidence through Monte Carlo sampling. We implemented this for the
priors and likelihood described above, and found that it works very
well for the problem under consideration.

\subsection{Maximum-likelihood analysis}

We also perform a standard frequentist maximum-likelihood analysis by
computing the maximum-likelihood modulation parameters for isotropic
Monte Carlo simulations. For these computations, we use a modified
version of the evidence code, which we find to be considerably more
robust than a simple non-linear search; while the non-linear search
algorithms often get trapped in local minima, the nested sampling
algorithm always find the correct solution, but of course, at a
considerably higher computational expense.

\section{Data}
\label{sec:data}

We analyze two versions of the three-year WMAP sky maps in the
following; the template-corrected $Q$-, $V$-, and $W$-band maps, and the
``foreground cleaned'' Internal Linear Combination (ILC) map
\citep{hinshaw:2007}. All maps are processed as described by
\citet{eriksen:2006}: They are first downgraded to
HEALPix\footnote{http://healpix.jpl.nasa.gov} resolution
$N_{\textrm{side}}=16$, by additional smoothing to a $9^{\circ}$ FWHM
Gaussian beam and appropriate pixel window. Second, uniform Gaussian
noise of $\sigma_{\textrm{n}} = 1\,\mu\textrm{K}$ is added to each
pixel in order to regularize the pixel-pixel covariance matrix. This
combination of smoothing and noise level results in a signal-to-noise
ratio of unity at $\ell = 40$, and strong noise domination at the
Nyquist multipole of $\ell = 47$.

\begin{figure}
\plotone{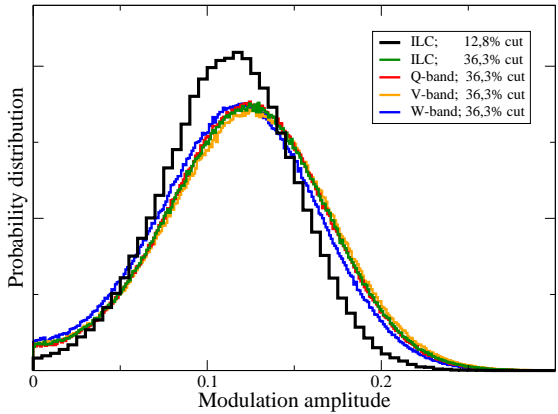}
\caption{Posterior distributions for the dipole modulation amplitude,
  marginalized over direction and CMB power spectrum.}
\label{fig:amplitude}
\end{figure}

We use two different sky cuts for our analyses. First, given that the
galactic plane is clearly visible in the single frequency data, our
first mask is conservatively defined.  This cut is created by
expanding the Kp2 mask \citet{hinshaw:2007} by $9^{\circ}$ in all
directions, and then manually removing all near-galactic pixels for
which any difference map between two channels are clearly larger than
noise. In total, 36.3\% of all pixels are rejected by this cut (see
figure \ref{fig:modulation_maps}). Second, we also adopt the directly
downgraded Kp2 cut used by the WMAP team that removes 12.8\% of all
pixels. We use this mask for the ILC map only.

The noise covariance matrix is given by the uniform noise only,
$N_{ij} = \sigma_{\textrm{n}}^2 \delta_{ij}$. For completeness, we
have also computed the noise covariance from the smoothed instrumental
noise for the $V$-band data, but we find that this has no effect on
the final results, since its amplitude is far below the CMB signal. It
is therefore omitted in the following.

\begin{deluxetable}{lccccc}
\tabletypesize{\small}
\tablecaption{\label{tab:results}}
\tablecomments{Modulation model results. The listed quantities are the
  marginal best-fit dipole axis (\emph{second column}) and amplitude
  (\emph{third column}); the change in likelihood at the posterior
  maximum, $\Delta \log \mathcal{L} = \log \mathcal{L}_{\textrm{mod}}
  -\log \mathcal{L}_{\textrm{iso}} $, between the modulated and the
  isotropic model (\emph{fourth column}); the Bayesian
  evidence difference, $\Delta \log E = \log E_{\textrm{mod}} - \log
  E_{\textrm{iso}}$ (\emph{fifth column}); and the
  frequentist probability for obtaining a lower maximum-likelihood
  modulation amplitude than the observed one, computed from isotropic
  simulations (\emph{sixth column}).}
\tablecolumns{6}
\tablehead{Data  & ($l_{\textrm{bf}}, b_{\textrm{bf}}$) &
  $A_{\textrm{bf}}$ & $\Delta \log \mathcal{L}$ & $\Delta \log E$ & $P$}
\startdata
ILC\tablenotemark{a}  & ($225^{\circ}, -27^{\circ}$)  & $0.114$ & 6.1 & $1.8\pm0.2$ & 0.991 \\
ILC\tablenotemark{b}  & ($208^{\circ}, -27^{\circ}$)  & $0.125$ & 6.0 & $1.8\pm0.2$  & 0.991\\
$Q$-band\tablenotemark{b} & ($222^{\circ}, -35^{\circ}$)  & $0.124$ & 5.5 & $1.5\pm0.2$ & 0.987 \\
$V$-band\tablenotemark{b} & ($205^{\circ}, -19^{\circ}$)  & $0.127$ & 5.6 & $1.5\pm0.2$ & 0.990 \\
$W$-band\tablenotemark{b} & ($204^{\circ}, -31^{\circ}$)  & $0.121$ & 5.2 & $1.3\pm0.2$ & $\,\,\,\,\,\,$0.985 

\enddata
\tablenotetext{a}{Liberal 12.8\% sky cut imposed.}
\tablenotetext{b}{Conservative 36.3\% sky cut imposed.}
\end{deluxetable}

As an additional hedge against foreground contamination, we
marginalize over a set of fixed spatial templates, $\mathbf{t}_i$,
through the covariance matrix $\mathbf{F}_i =
\alpha_i\mathbf{t}_i\mathbf{t}_i^T$, $\alpha_i \gtrsim 10^3$. Monopole
and dipole terms are always included, and one or more foreground
templates. For the $V$-band and ILC maps, we follow
\citet{hinshaw:2007} and adopt $V$--ILC as our foreground template.
For the $Q$-band data, we marginalize over a synchrotron
\citep{haslam:1982}, a free-free \citep{finkbeiner:2003}, and a dust
\citep{finkbeiner:1999} template individually. Finally, for the
$W$-band data, we use the $W$--ILC difference map. However, we have
tried various combinations for all maps, and there is virtually no
sensitivity to the particular choice, or indeed, to the template at
all, due to the conservative sky cut used.

\section{Results}
\label{sec:results}

\begin{figure}
\plotone{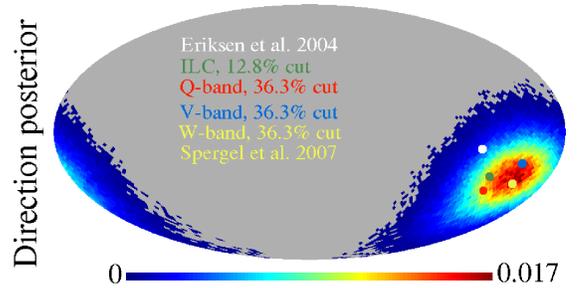}
\caption{Posterior distribution for the dipole modulation
  axis, shown for the ILC map and 36.3\% sky cut, and marginalized
  over power spectrum and amplitude parameters. Grey sky pixels
  indicate pixels outside the $2\sigma$ confidence region. The dots
  indicate the axis 1) reported by Eriksen et al.\ (2004) in white; 2)
  for the ILC map with a 12.8\% sky cut in green; 3) for the $Q$-, $V$-,
  and $W$-bands in red, blue and yellow, respectively. The axis reported
  by \citet{spergel:2007a} coincides with the $W$-band axis.}
\label{fig:modulation_maps}
\end{figure}

The results from the analysis outlined above are summarized in Table
\ref{tab:results}. For each map, we report the best-fit dipole axis
and amplitude, as well as the maximum log-likelihood difference and
Bayesian evidence difference for the modulated versus the isotropic
model. The errors on the evidence are estimated by performing eight
independent analyses for each case, and computing the standard
deviation \citep{mukherjee:2006}. We also compute the probability of
obtaining a smaller modulation amplitude than the observed one by
analyzing 1000 isotropic Monte Carlo simulations.

Starting with the first case in Table \ref{tab:results}, the ILC map
cut by a 12.8\% mask, we see that the best-fit modulation axis points
toward $(l,b) = (225^{\circ}, -27^{\circ})$, and the corresponding
modulation amplitude is 0.114.  The raw likelihood improvement is
$\Delta \log \mathcal{L} = 6.1$. The probability of finding such a
high modulation amplitude in intrinsically isotropic simulations is
$\sim1$\%, and, finally, the improvement in Bayesian evidence is
$\Delta \log E = 1.8$. 

Further, these results are not sensitive to data set or sky coverage:
Even the Q-band map, which presumably is the least reliable with
respect to residual foregrounds, yields a modulation amplitude which
is high at the 98.7\% (frequentist) confidence level, and a Bayesian
log-evidence improvement of 1.5. This frequency independence is
further illustrated in Figure \ref{fig:amplitude}, where we show the
marginalized posterior distributions for the modulation amplitudes for
each data set. The agreement among data sets is very good.

In Figure \ref{fig:modulation_maps} we show the dipole axis posterior
distribution for the ILC map and 36.3\% sky cut.  Superimposed on
this, we have also marked the first-year asymmetry axis reported by
\citet{eriksen:2004b} [$(l,b) = (237^{\circ}, -10^{\circ})$] in
yellow, and also the other axes listed in Table \ref{tab:results}. All
agree well within $2\sigma$, and this is another testimony to the
excellent stability of the effect with respect to statistical method,
data set and overall procedure.

Finally, we note that this model may also partially explain the
anomalous cold spot reported by \citet{vielva:2004} and
\citet{cruz:2005}: By demodulation the spot would increase its
temperature by about 10\%, and although still very cold, it would be
significantly less extreme. Similar arguments could possibly also be
made for the Bianchi VII$_h$ correlation found by \citet{jaffe:2005}.
These issues will be considered further in future work.

\section{Conclusions}
\label{sec:conclusions}

A notable power asymmetry between two opposing hemispheres in the
first-year WMAP sky maps was reported by \citet{eriksen:2004b}. This
feature may be observed as strong fluctuations in the southern
ecliptic hemisphere, but virtually no large-scale structure in the
northern ecliptic hemisphere (e.g., Hinshaw et al.\ 2007).

In this Letter, we have revisited this issue in the three-year WMAP
data, adopting the statistical framework introduced and applied by
\citet{spergel:2007a}. With these tools, we find that the evidence for
power asymmetry in the WMAP data is very consistent with that
initially reported for the first-year maps by \citet{eriksen:2004b},
and the WMAP data clearly suggest a dipolar distribution of power on
the sky: The best-fit modulation amplitude is roughly 12\% in real
space, or about 20\% in terms of power spectra.  The corresponding
dipole direction is $(l,b) \sim (225^{\circ}, -27^{\circ})$. All
results are independent of data set choices, i.e., frequency channel
or sky cut.

However, the statistical evidence for this effect is still only
tentative. In frequentist language, the significance is about 99\%,
while in Bayesian terms, the log-evidence difference is $\sim 1.5$ to
1.8, corresponding to odds of one to five or six. This
is quite comparable to the evidence for $n_{\textrm{s}} \ne 1$ after
the three-year WMAP data release, for which the odds are about one to
eight in the highest case (Parkinson et al.\ 2006).  Thus, there is
still a chance that the effect may be a fluke, and most likely, this
will remain the situation until Planck provides new data in some five
years. With additional frequency coverage, a better job can be done on
foreground treatment, and more sky coverage can be reliably included
in the analysis. Second, full-sky and high-sensitivity polarization
data should provide valuable insights on the origin of the effect.

\begin{acknowledgements}
  HKE acknowledges financial support from the Research Council of
  Norway.  Some of the results in this paper have been derived using
  the HEALPix (G\'orski et al.\ 1999) software and analysis package.
  We acknowledge use of the Legacy Archive for Microwave Background
  Data Analysis (LAMBDA). Support for LAMBDA is provided by the NASA
  Office of Space Science.  
\end{acknowledgements}

\end{document}